\documentclass[11pt,letterpaper]{article}
\usepackage{jheppub}

\usepackage{graphicx}

\usepackage{amsmath}
\usepackage{subfigure}
\usepackage{url}

\usepackage{multirow}
\usepackage{threeparttable}
\usepackage{paralist}

\newcommand{\MPl}{M_{\text{Pl}}}
\newcommand{\NL}{\text{NL}}

\newcommand{\bPhi}{{\boldsymbol \Phi}}

\newcommand{\bs}[1]{\boldsymbol #1}

\newcommand{\bC}{{\boldsymbol C}}
\newcommand{\bZ}{{\boldsymbol Z}}
\newcommand{\bK}{{\boldsymbol K}}
\newcommand{\bX}{{\boldsymbol X}}

\newcommand{\bQ}{{\boldsymbol Q}}
\newcommand{\boldf}{{\boldsymbol f}}
\newcommand{\bOmega}{{\boldsymbol \Omega}}
\newcommand{\bW}{{\boldsymbol W}}
\newcommand{\bR}{{\boldsymbol R}}

\newcommand{\vev}[1]{\left\langle #1 \right\rangle}

\newcommand{\gold}{\widetilde{G}_L}
\newcommand{\goldbar}{\overline{\widetilde{G}}_L}

\newcommand{\Arg}{\mathop{\text{Arg}}}

\DeclareRobustCommand{\Sec}[1]{Sec.~\ref{#1}}

\DeclareRobustCommand{\App}[1]{App.~\ref{#1}}
\DeclareRobustCommand{\Tab}[1]{Table~\ref{#1}}

\DeclareRobustCommand{\Eq}[1]{Eq.~(\ref{#1})}
\DeclareRobustCommand{\Eqs}[2]{Eqs.~(\ref{#1}) and (\ref{#2})}
\DeclareRobustCommand{\Ref}[1]{Ref.~\cite{#1}}
\DeclareRobustCommand{\Refs}[1]{Refs.~\cite{#1}}

\newcommand{\be}{\begin{equation}}
\newcommand{\ee}{\end{equation}}

\renewcommand{\i}{{\bar{\imath}}}
\renewcommand{\j}{{\bar{\jmath}}}

\newcommand{\D}{{\mathcal{D}}}
\newcommand{\sigmabar}{\overline{\sigma}}

\title{The Two Faces of Anomaly Mediation}

\author{Francesco D'Eramo,}
\author{Jesse Thaler,}
\author{and Zachary Thomas}

\emailAdd{fderamo@mit.edu}
\emailAdd{jthaler@mit.edu}
\emailAdd{ztt@mit.edu}

\affiliation{Center for Theoretical Physics, Massachusetts Institute of Technology, Cambridge, MA 02139, USA}

\abstract{Anomaly mediation is a ubiquitous source of supersymmetry (SUSY) breaking which appears in almost every theory of supergravity.  In this paper, we show that anomaly mediation really consists of two physically distinct phenomena, which we dub ``gravitino mediation'' and ``K\"ahler mediation''.  Gravitino mediation arises from minimally uplifting SUSY anti-de Sitter (AdS) space to Minkowski space, generating soft masses proportional to the gravitino mass.  K\"ahler mediation arises when visible sector fields have linear couplings to SUSY breaking in the K\"ahler potential, generating soft masses proportional to beta function coefficients.  In the literature, these two phenomena are lumped together under the name ``anomaly mediation'', but here we demonstrate that they can be physically disentangled by measuring associated couplings to the goldstino.  In particular, we use the example of gaugino soft masses to show that gravitino mediation generates soft masses without corresponding goldstino couplings.  This result naively violates the goldstino equivalence theorem but is in fact necessary for supercurrent conservation in AdS space.  Since gravitino mediation persists even when the visible sector is sequestered from SUSY breaking, we can use the absence of goldstino couplings as an unambiguous definition of sequestering.}


\begin{document}

\hfill MIT-CTP 4345

\maketitle

\section{Introduction}
\label{sec:introduction}

If supersymmetry (SUSY) is realized in nature, then SUSY must be spontaneously broken and the effects of SUSY breaking must be mediated to the supersymmetric standard model (SSM).  In the context of supergravity (SUGRA), the most ubiquitous form of mediation is ``anomaly mediation'' \cite{Randall:1998uk,Giudice:1998xp,Pomarol:1999ie,Gherghetta:1999sw}, which persists even when (and especially when) a SUSY-breaking hidden sector is sequestered from the visible sector.  Of course, anomaly mediation need not be the dominant source of SSM soft masses, and there are theories where anomaly mediation is suppressed or absent \cite{Lahanas:1986uc,Luty:2002hj,Luty:2002ff}.  But given its ubiquity, it is worth better understanding the physics of anomaly mediation and the circumstances which give rise to sequestering.

Indeed, anomaly mediation has been the source of much theoretical confusion, and various papers have aimed to clarify the underlying mechanism \cite{Chacko:1999am,Bagger:1999rd,Bagger:2000dh,Dine:2007me,deAlwis:2008aq,Jung:2009dg,Conlon:2010qy,Sanford:2010hc}.  The original description of anomaly mediation involved the super-Weyl anomaly \cite{Randall:1998uk,Giudice:1998xp}, and the most straightforward derivation of anomaly-mediated soft masses uses the conformal compensator formalism of SUGRA \cite{Gates:1983nr}.  As discussed in \Ref{Bagger:1999rd}, anomaly mediation really involves three different anomalies:  a super-Weyl anomaly, a K\"ahler anomaly, and a sigma-model anomaly.  More recently, \Ref{Dine:2007me} emphasized that SUGRA is not even a necessary ingredient, as a version of anomaly mediation (corresponding to the sigma-model anomaly) appears even in the $\MPl \rightarrow \infty$ limit.\footnote{\Ref{Dine:2007me} also emphasized that the language of ``anomalies'' is not necessary, as the effect can be alternatively described in terms of gaugino counterterms.  These gaugino counterterms are necessary to maintain SUSY in the 1PI effective action, including all anomaly contributions.}

In this paper, we will show that the phenomenon known as ``anomaly mediation'' really consists of two physically distinct effects.  This realization clarifies a number of confusions surrounding anomaly mediation, and leads to a physical definition of sequestering in terms of goldstino couplings. Throughout this paper, we will use ``goldstino'' to refer to the longitudinal gravitino mode in the high energy limit ($E \gg m_{3/2}$) \cite{Fayet:1977vd,Casalbuoni:1988qd, Casalbuoni:1988kv}.\footnote{For $\MPl \rightarrow \infty$ and $m_{3/2} \rightarrow 0$, this mode is the true goldstino from spontaneous SUSY breaking \cite{Volkov:1972jx,Rocek:1978nb,Samuel:1982uh}.   Here, we will keep $m_{3/2}$ fixed by considering the goldstino mode in rigid AdS space \cite{Keck:1974se,Zumino:1977av,Ivanov:1979ft,Ivanov:1980vb}.  In particular, the familiar relation $m_{3/2} = F_{\rm eff}/\sqrt{3}\MPl$ is only true after adjusting the cosmological constant to zero, so we can still take $\MPl \rightarrow \infty$ while preserving effects proportional to $m_{3/2}/F_{\rm eff}$.}  The two effects are as follows.
\begin{itemize}
\item \textbf{Gravitino Mediation}.  Famously, bosonic and fermionic modes in the same multiplet have SUSY mass splittings in the bulk of four-dimensional anti-de Sitter (AdS) space \cite{Breitenlohner:1982jf,Nicolai:1984hb,Gripaios:2008rg}.\footnote{These splittings are required by the global AdS SUSY algebra.  The case of massless gauge multiplets is subtle, since physical gauginos are massless in AdS$_4$.  Crucially, a bulk gaugino mass term is required to cancel an infrared contribution to the gaugino mass in the 1-loop 1PI effective action, arising from boundary conditions in AdS$_4$ \cite{Gripaios:2008rg}.} These mass splittings are proportional to the AdS curvature, and thus to the gravitino mass $m_{3/2}$.  If SUSY AdS space is minimally uplifted to Minkowski space via SUSY breaking, these mass splittings are preserved, leading to SSM soft masses from ``gravitino mediation''.  These soft masses do not have associated couplings to the goldstino, naively violating the (flat space) goldstino equivalence theorem \cite{Fayet:1977vd,Casalbuoni:1988qd, Casalbuoni:1988kv}.  Nevertheless, the absence of goldstino couplings is necessary for conservation of the AdS$_4$ supercurrent.  Gravitino mediation closely resembles traditional anomaly mediation \cite{Randall:1998uk,Giudice:1998xp}, and is related to the super-Weyl anomaly.  Gravitino mediation can never be turned off, since it arises from the infrared symmetry structure of SUSY AdS space.
\item \textbf{K\"ahler Mediation}.   If visible sector fields have linear couplings to SUSY-breaking fields in the K\"ahler potential, then this gives rise to ``K\"ahler mediation'', where SSM fields get mass splittings proportional to beta function coefficients.  Linear couplings are ubiquitous in the presence of modulus fields, so K\"ahler mediation typically accompanies (and sometimes cancels) gravitino mediation in explicit SUGRA constructions \cite{Lahanas:1986uc,Luty:2002hj,Luty:2002ff,ArkaniHamed:2004yi}.   As expected from the (flat space) goldstino equivalence principle, these soft masses have a corresponding coupling to the goldstino.   In the $\MPl \rightarrow \infty$ limit, K\"ahler mediation appears via the sigma-model anomaly (as emphasized in \Ref{Dine:2007me}).  It also receives $1/\MPl$ corrections due to the super-Weyl and K\"ahler anomalies.  Unlike gravitino mediation, K\"ahler mediation is sensitive to the ultraviolet couplings of the theory.
\end{itemize}
These two contributions to anomaly mediation are summarized in \Tab{tab:summary}, focusing on the case of gaugino soft masses.  Full anomaly mediation is simply the sum of gravitino mediation and K\"ahler mediation.\footnote{\label{foot:conlon}As pointed out in \Ref{Conlon:2010qy} in the context of string theory, there is an additional anomaly-mediated gaugino mass which arises from an anomalous rescaling of the gauge multiplets.  We discuss this effect in \App{app:fullmass} and show that it yields a corresponding goldstino coupling consistent with (flat space) supercurrent conservation.}

\renewcommand{\arraystretch}{1.5}
\begin{table}
\begin{center}
\begin{tabular}{|c||cccc|}
\hline
& Anomaly? & $m_\lambda \propto $ ? & SUGRA? & Goldstino? \\
\hline \hline
\textbf{Gravitino Mediation} & Super-Weyl & $(3 T_G - T_R) m_{3/2}$ & \checkmark & \\
\hline \hline
\multirow{3}{*}{\textbf{K\"ahler Mediation}} & Super-Weyl &$\frac{1}{3}(3 T_G - T_R) K_i F^i$ & \checkmark & \checkmark \\
 & K\"ahler &  $-\frac{2}{3}T_R K_i F^i$ & \checkmark & \checkmark \\
 & Sigma-Model & $2 \frac{T_R}{d_R} (\log \det K|_R'')_iF^i$ &  & \checkmark\\
 \hline
\end{tabular}
\end{center}
\caption{The two faces of anomaly mediation.  Shown are the corresponding anomalies and their contributions to gaugino masses, with a notation to be explained in the body of the text.  (All the masses have an overall factor of $-g^2/16\pi^2$.)  We indicate whether the effect requires SUGRA and whether there is an associated gauge boson-gaugino-goldstino coupling.  Gravitino mediation can be distinguished from K\"ahler mediation by the goldstino coupling.  If SUSY breaking couples directly to gauginos, then there is an additional anomaly contribution discussed in \App{app:fullmass}, which yields both one-loop gaugino masses and goldstino couplings.}
\label{tab:summary}
\end{table}

One might naively expect that no physical measurement could distinguish between gravitino mediation and K\"ahler mediation, since they only appear in combination in SSM soft masses.  However, there is a crucial physical distinction in terms of goldstino couplings.\footnote{Our results can be interpreted as describing goldstino couplings in the analog of Landau gauge where the gravitino field is purely transverse.  At the end of \Sec{sec:goldstinoequivalence}, we explain the same effect in unitary gauge.}  In usual SUSY breaking scenarios, gaugino soft masses are accompanied by a corresponding coupling between the gaugino $\lambda^a$, the gauge boson $A_{\mu}^a$, and the goldstino $\gold$,
\be
\label{eq:genericGoldstinoCouplings}
\mathcal{L} \supset - \frac{1}{2}m_{\lambda} \lambda_a \lambda^a + \frac{c_\lambda}{\sqrt{2} F_{\rm eff}} \lambda_a \sigma^{\mu \nu} \gold F_{\mu \nu}^a,
\ee
where $F_{\rm eff}$ is the scale of SUSY breaking.\footnote{There is also an additional coupling between the gaugino, goldstino, and the auxiliary field $D^a$, $\frac{ i c_\lambda}{\sqrt{2} F_{\rm eff}} \lambda_a \gold D^a$.  The $c_\lambda$ of this coupling is guaranteed to be identical to the $c_\lambda$ in \Eq{eq:genericGoldstinoCouplings}, so we omit this term for brevity throughout.}  For global SUSY, the relation $c_\lambda = m_{\lambda}$ is required by the (flat space) goldstino equivalence theorem.   In contrast, gravitino mediation is dictated by SUSY in AdS space, and generates a contribution to $m_{\lambda}$ \emph{without} a corresponding contribution to $c_\lambda$.  Indeed, the difference $m_{\lambda} - c_\lambda$ is necessarily proportional to $m_{3/2}$ by conservation of the AdS supercurrent, and this gives a physical way to measure gravitino mediation as distinct from all other sources of SSM soft masses.  We will show this explicitly in \Eq{eq:finalAnswer}.

This result allows us to give an unambiguous definition of sequestering \cite{Randall:1998uk}, which is the condition necessary for traditional anomaly mediation (i.e.~gravitino mediation) to be the sole source of SSM soft masses.
\begin{itemize}
\item Visible sector fields are sequestered from SUSY breaking if they do not have linear couplings to the goldstino.\footnote{Strictly speaking this is only true for gauginos.  As we will explain below, scalar soft masses are more subtle because of irreducible couplings to the goldstino, but sequestering for scalars can still be defined as the absence of any further couplings to the goldstino.}
\end{itemize}
In other words, $c_\lambda$ is measure of how well the visible sector is sequestered from the goldstino.  Previously, sequestering was known to occur when the K\"ahler potential $K$ and superpotential $W$ took a special ``factorized'' form \cite{Randall:1998uk}
\be
\label{eq:factorized_form}
-3 e^{- K/3} = \Omega_{\rm vis} + \Omega_{\rm hid}, \qquad W = W_{\rm vis} + W_{\rm hid}.
\ee
However, \Eq{eq:factorized_form} is ambiguous, since the separation into ``visible'' and ``hidden'' sectors is not robust to K\"ahler transformations by a chiral multiplet $X$ with $K \rightarrow K + X + X^\dagger$ and $W \rightarrow e^{-X} W$.   Also, sequestering usually (but not always) requires moduli to be stabilized \cite{Goldberger:1999uk,Luty:1999cz,ArkaniHamed:1999pv,Luty:2000ec,Maru:2003mq,Eto:2004yk}.  Sequestering does have an unambiguous extra-dimensional interpretation in terms of geometric separation \cite{Randall:1998uk}.  Here we can use the absence of goldstino couplings as a purely four-dimensional definition of sequestering.  Since physical couplings are invariant to Lagrangian manipulations such as K\"ahler transformations, this definition does not suffer from the ambiguities of \Eq{eq:factorized_form}, and gives a robust criteria for determining when traditional anomaly mediation is dominant.

We can highlight the distinction between gravitino mediation alone and anomaly mediation more generally by comparing models with strict sequestering \cite{Randall:1998uk} to models with warped \cite{Luty:1999cz,ArkaniHamed:1999pv,Luty:2000ec,Luty:2002ff,Maru:2003mq,Eto:2004yk,Son:2008mk} or conformal sequestering \cite{Nelson:2000sn,Kobayashi:2001kz,Nelson:2001mq,Luty:2001jh,Kobayashi:2001is,Luty:2001zv,Kobayashi:2002iz,Dine:2004dv,Ibe:2005pj,Ibe:2005qv,Schmaltz:2006qs}. In the case of strict sequestering, SUSY breaking is confined to a brane which is geometrically separated from the visible sector brane.  This geometric separation \emph{forbids} couplings between the goldstino and the visible sector.  The only source of visible sector soft masses comes from gravitino mediation, which can be captured by the conformal compensator\footnote{The relation $\vev{F_\Phi} = m_{3/2}$ is special to strict sequestering.  See \Eq{eq:improvedGauge} for a more general expression.}
\be
\vev{\bPhi} = 1 + \theta^2 m_{3/2}.
\ee
In the case of warped sequestering, visible sector fields on the IR brane feel an ``effective'' conformal compensator
\be
{\bs \omega} = \bPhi e^{-k {\bs T}},
\ee
where ${\bs T}$ is the radion superfield.  Visible sector fields obtain anomaly-mediatied soft masses proportional to
\be
\vev{\frac{F_\omega}{\omega}} = m_{3/2}  - k F_T,
\ee
but because the radion has overlap with the goldstino direction, there are visible sector couplings to the goldstino proportional to $k F_T$.  In the language of this paper, warped sequestering exhibits a cancellation between gravitino mediation and K\"ahler mediation.\footnote{This cancellation is not a fine tuning, since it arises from the geometry of the warped (AdS$_5$) space.  The curvature of AdS$_5$ should not be confused with the curvature of AdS$_4$, which is responsible for gravitino mediation.}

Throughout this paper, we focus on gaugino masses, leaving a full description of anomaly-mediated scalar soft masses to future work \cite{scalars:2012}.  As a preview, there is a mass splitting between scalars and matter fermions in the bulk of AdS$_4$, analogous to the gaugino case, which includes the familiar two-loop anomaly-mediated scalar masses.  However, already at tree-level in AdS$_4$, scalars have tachyonic scalar soft masses equal to $-2 m_{3/2}^2$ \cite{Breitenlohner:1982jf,Nicolai:1984hb}.  While tachyonic scalar masses do not destabilize the theory in AdS space, they do in flat space.  Thus, the SUSY breaking that uplifts the theory from AdS to flat space must remove these tree-level tachyonic soft masses, resulting in irreducible goldstino couplings which complicate the definition of sequestering.\footnote{There is a related subtlety involving tree-level holomorphic $B$-terms, since $B$ terms arising from the superpotential have different associated goldstino couplings than $B$ terms arising from the Giudice-Masiero mechanism \cite{Giudice:1988yz}.  Previously, both phenomena were considered to occur in the sequestered limit, but Giudice-Masiero secretly violates the conditions for sequestering \cite{scalars:2012}.}

In the remainder of this paper, we derive the gaugino soft masses and goldstino couplings arising from anomaly mediation, emphasizing the distinction between gravitino mediation and K\"ahler mediation.  The soft masses are well-known in the literature, but to the best of our knowledge, the goldstino couplings have never been calculated explicitly.  In \Sec{sec:anomalyInGlobal}, we give a straightforward derivation of how K\"ahler mediation arises in global SUSY.  We then turn to full SUGRA in \Sec{sec:gaugefix}, applying the improved SUGRA gauge fixing of \Ref{Cheung:2011jp}.  This is the simplest way to isolate gravitino mediation, since this gauge automatically decouples the (transverse) gravitino, leaving the goldstino coupling manifest.  In \Sec{sec:SUGRAmultiplet}, we describe the same physics using a more conventional SUGRA notation of \Ref{Bagger:1999rd}.  We also explain the connection to the AdS$_4$ supercurrent conservation and the goldstino equivalence theorem.  We conclude in \Sec{sec:conclusions}.   

\section{K\"ahler Mediation in Global SUSY}
\label{sec:anomalyInGlobal}

Before deriving full anomaly mediation in \Sec{sec:gaugefix}, it is useful to focus on the case of pure K\"ahler mediation, which arises in the limit of global SUSY.  This example was emphasized in \Ref{Dine:2007me}, but in order to connect to the (perhaps) more familiar language of \Ref{Bagger:1999rd}, we will derive the result using chiral anomalies (instead of gaugino counterterms).

Consider a field redefinition acting on a chiral superfield $\bQ$ of the form
\be
{\bs Q} \rightarrow e^{{\bs \alpha}} {\bs Q},
\ee
where ${\bs \alpha}$ is another chiral superfield.\footnote{Throughout this paper, we will use the notation of \Ref{Cheung:2011jp}, where boldface ($\bX$) indicates a superfield and regular typeface ($X$) indicates the lowest component of the corresponding superfield.  Superscripts are field labels and subscripts indicate derivatives with respect to chiral fields.  As needed, we raise and lower indices using the K\"ahler metric.  We will use $\bQ$ to indicate visible sector fields and $\bX$ to indicate hidden sector SUSY-breaking fields.}  This field redefinition changes the Lagrangian in a classical way, but it also introduces a term related to the Konishi anomaly \cite{Clark:1979te,Konishi:1983hf}.  If $\bQ$ is in the representation $R$ of non-Abelian gauge field, then the Lagrangian shifts as
\be
\label{eq:SUSYshift}
\mathcal{L} (\bX) \rightarrow \mathcal{L}(e^{{\bs \alpha}} {\bs X}) + \frac{g^2\, T_R}{16\pi^2} \int d^2 \theta  \, {\bs \alpha}   {\bW}^{a \alpha} {\bW}^a_\alpha,
\ee
where $T_R$ is the Dynkin index of the representation $R$.  In the language of \Ref{Giudice:1997ni,ArkaniHamed:1998kj}, \Eq{eq:SUSYshift} is simply the chiral anomaly analytically continued into superspace.

In global SUSY, K\"ahler mediation arises whenever charged matter has linear couplings to SUSY breaking in the K\"ahler potential.  This is easiest to understand using a non-linear representation $\bX_{\NL}$ of a SUSY-breaking field which obeys the constraint $\bX_{\NL}^2 = 0$ \cite{Rocek:1978nb,Lindstrom:1979kq,Komargodski:2009rz,Cheung:2010mc,Cheung:2011jq}.  Consider a Lagrangian which contains a matter field $\bQ$ coupled to SUSY breaking as
\be
\mathcal{L} \supset \int d^4 \theta \, \bQ^\dagger \bQ \left(1 + \frac{\bX_{\NL} + \bX^\dagger_{\NL}}{\Lambda}  \right).
\ee
We can remove the linear couplings of $\bX_{\NL}$ by performing an (anomalous) field redefinition
\be
\bQ \rightarrow \bQ \left(1- \frac{\bX_{\NL}}{\Lambda} \right) = \bQ e^{- \bX_{\NL}/ \Lambda}, \label{eq:simplefieldredef}
\ee
where the last equality relies on $\bX_{\NL}^2 = 0$.  From the Konishi anomaly, this yields
\be
\label{eq:nonlinearSUSYshift}
\mathcal{L} \supset \int d^4 \theta \, \bQ^\dagger \bQ \left(1 - \frac{\bX_{\NL} \bX^\dagger_{\NL}}{\Lambda^2} \right) - \frac{g^2\, T_R}{16\pi^2} \int d^2 \theta  \, \frac{\bX_{\NL}}{\Lambda}   {\bW}^{a \alpha} {\bW}^a_\alpha,
\ee
After the field redefinition, $\bX_{\NL}$ only has quadratic couplings to $\bQ$, at the expense of introducing new couplings between $\bX_{\NL}$ and the gauge multiplet.  This is the essence of K\"ahler mediation.

Expanding out $\bX_{\NL}$ in terms of $F_X$ and the goldstino $\gold$ \cite{Rocek:1978nb,Lindstrom:1979kq,Komargodski:2009rz,Cheung:2010mc,Cheung:2011jq}
\be
\bX_{\NL} = \left(\theta + \frac{1}{\sqrt{2}} \frac{\gold}{F_X} \right)^2 F_X,
\ee
\Eq{eq:nonlinearSUSYshift} contains a soft mass for the gauginos and a corresponding coupling to the goldstino, as anticipated in \Eq{eq:genericGoldstinoCouplings}
\be
\tag{\ref{eq:genericGoldstinoCouplings}}
\mathcal{L} \supset - \frac{1}{2}m_{\lambda} \lambda_a \lambda^a + \frac{c_\lambda}{\sqrt{2} F_{\rm eff}} \lambda_a \sigma^{\mu \nu} \gold F_{\mu \nu}^a,
\ee
where $F_{\rm eff} \equiv F_X$ in this example, and
\be
m_\lambda = c_\lambda =  - \frac{g^2 \, T_R}{8 \pi^2}  \frac{F_X}{\Lambda}. \label{eq:simplegauginomass}
\ee
As expected from the goldstino equivalence theorem (see \Sec{sec:goldstinoequivalence}), the goldstino couplings are proportional to the gaugino mass.

In the above derivation, the matter superfield $\bQ$ was assumed to be massless, which was crucial for seeing a physical effect from the sigma-model anomaly.  Indeed, without massless ``messengers'' to communicate SUSY breaking, one does not expect K\"ahler potential terms to affect holomorphic quantities like gaugino masses.  To see what happens for massive matter, consider vector-like chiral superfields with a supersymmetric mass term $\mu \bQ \bQ^c$.  In this case, the chiral rescaling in \Eq{eq:simplefieldredef} yields a new superpotential term $- \frac{\mu}{\Lambda} \bQ \bQ^c \bX_{\NL}$.  For large $\mu$, $\bQ$ and $\bQ^c$ are just heavy messenger fields, generating a gauge-mediated contribution to the gaugino masses which exactly cancels \Eq{eq:simplegauginomass}, as explicitly shown in \Refs{Seiberg:2008qj,Dumitrescu:2010ha}.  This insensitivity to heavy supersymmetric thresholds is a well-known feature of anomaly mediation, and persists in SUGRA as well; we may in general evaluate anomaly or beta-function coefficients at the scale of interest.  For simplicity, we will take all matter superfields to be massless in the remainder of this paper.

The chiral rescaling procedure in \Eq{eq:simplefieldredef} can be generalized to an arbitrary K\"ahler potential $\bK$.  
\be
\label{eq:globalKahler}
\mathcal{L} \supset \int d^4 \theta \bK.
\ee
Consider a set of chiral multiplets $\bQ$ in the representation $R$ with the K\"ahler metric $\bK|_R''$.\footnote{The K\"ahler metric $\bK|_R''$ is just $\bK_{i\j}$ where $\bQ^i$ and $\bQ^j$ transform in $R$.}  In general, $\bK|_R''$ will be a function of SUSY-breaking fields $\bX^i$, but as shown in \App{app:generalKahler}, there is a field redefinition that removes all linear couplings of $\bX^i$ in $\bK|_R''$ but generates the anomalous term
\be
\label{eq:generalRedef}
\delta \mathcal{L} = - \int d^2 \theta \frac{g^2}{16\pi^2}  {\bW}^{a \alpha} {\bW}^a_\alpha   \frac{\overline{D}^2 D^2}{16 \, \square} \left[  \frac{T_R}{d_R} \log \det \bK|_R'' \right],
\ee
where $d_R$ is dimension of the representation $R$.  This form makes explicit use of the chiral projection operator ($\overline{D}^2 D^2/16 \,\square$), which is overkill for our purposes.  Since we are only interested in soft masses and goldstino couplings, we will assume that all SUSY-breaking fields have been shifted such that $\vev{X^i} = 0$, and focus on a subset of terms from expanding \Eq{eq:generalRedef} to first order in $\bX^i$:
\be
\label{eq:generalRedefSub}
\delta \mathcal{L} \supset - \int d^2 \theta \frac{g^2}{16\pi^2}  \frac{T_R}{d_R} (\log \det K|_R'')_i \bX^i  {\bW}^{a \alpha} {\bW}^a_\alpha.
\ee
In each SUSY-breaking multiplet $\bX^i$, the fermionic component $\chi^i$ has overlap with the goldstino direction as
\be
\label{eq:goldstinoMode}
\chi^i \rightarrow \frac{F^i}{F_{\rm eff}} \gold,
\ee
where $F_{\rm eff}$ is the total amount of SUSY breaking (in the absence of $D$ terms, $F_{\rm eff} = \sqrt{F_i F^i}$).  We see that \Eq{eq:generalRedefSub} contains a gaugino mass and corresponding goldstino coupling
\be
m_\lambda = c_\lambda =  - \frac{g^2}{16\pi^2} \frac{2 T_R}{d_R} (\log \det K|_R'')_i F^i.
\ee
Once we sum over representations $R$, this is the general expression for K\"ahler mediation in global SUSY.  As we will see, this chiral field scaling procedure will persist when we go to SUGRA, but the equality between  $m_\lambda$ and $c_\lambda$ will be broken.

\section{Gravitino and K\"ahler Mediation in SUGRA}
\label{sec:gaugefix}

Having derived K\"ahler mediation in global SUSY, we can now understand the analogous effect in full SUGRA.  Now, the goldstino is eaten by the super-Higgs mechanism to become the longitudinal component of the gravitino, but it is still convenient to isolate the goldstino mode by using goldstino equivalence in the high energy limit.  For simplicity, we will use ``anomaly mediation'' to refer to the combined effect of gravitino and K\"ahler mediation.  As we will see, these two effects are physically distinct from the perspective of goldstino couplings.  

The improved SUGRA gauge fixing of \Ref{Cheung:2011jp} is particularly convenient for understanding anomaly mediation, both in terms of soft masses and goldstino couplings.  In this gauge, matter multiplets (including the goldstino multiplet) are decoupled from the gravity multiplet up to $1/\MPl^2$ suppressed effects.  This allows calculations involving the matter fields alone to be performed in \emph{global} superspace.  After giving a brief description of the SUGRA Lagrangian and the gauge fixing of \Ref{Cheung:2011jp}, we will calculate gaugino masses and goldstino couplings to see the two faces of anomaly mediation.

\subsection{The SUGRA Lagrangian}

The conformal compensator formalism arises from gauge fixing conformal SUGRA using the conformal compensator field $\bPhi$.  As reviewed in \Ref{Cheung:2011jp}, the tree-level SUGRA Lagrangian can be written as
\be
\label{eq:SUGRAlagrange}
\mathcal{L} = -3 \int d^4 \theta\, \bPhi^\dagger \bPhi e^{- \bK/3} + \int d^2 \theta \, \bPhi^3 \, \bW + \text{h.c.} + \frac{1}{4} \int d^2 \theta \, \boldf_{ab}
\bW^{a \alpha} \bW^b_\alpha + \text{h.c.} + \ldots,
\ee
where the ellipsis ($\ldots$) corresponds to terms involving the graviton and gravitino.  In general, the ellipsis contains quadratic mixing terms between matter multiplets and the graviton multiplet, but \Ref{Cheung:2011jp} showed that there is an improved gauge fixing for $\bPhi$ where this mixing is absent:
\begin{align}
\bPhi &= e^{\bZ/3} (1 + \theta^2 F_\phi), \label{eq:improvedGauge}\\
\bZ &= \vev{K/2 - i \Arg W} + \vev{K_i} \bX^i. 
\end{align}
In this gauge, one can simply drop the ellipsis terms in \Eq{eq:SUGRAlagrange} for any calculation not involving gravitons or gravitinos, allowing one to study matter multiplets in SUGRA using global superspace manipulations.

There are a few important caveats to this gauge fixing.  First, \Eq{eq:improvedGauge} only removes mixing terms at tree level, so strictly speaking, one can only study tree-level and one-loop effects using this formalism.  This is sufficient for understanding anomaly-mediated gaugino masses at one loop, but we will have to postpone a study of two-loop scalar soft masses for future work.  Second, this gauge fixing assumes that the cosmological constant has been adjusted to zero to yield a Minkowski vacuum, a necessary assumption for phenomenology.  Third, \Eq{eq:improvedGauge} explicitly contains vacuum expectation values (vevs), which is perhaps unfamiliar but conceptually sound.

A nice feature of this gauge is that after adjusting the cosmological constant to zero
\be
\label{eq:Fphivev}
\vev{F_\phi} = m_{3/2},
\ee
making it easy to identify terms proportional to the gravitino mass \cite{Cheung:2011jp}.  In particular, note that the $(1 + \theta^2 m_{3/2})$ part of $\bPhi$ has a SUSY-breaking $F$-component \emph{without} any coupling to fermions.  This will be the origin of gravitino mediation, which yields soft masses proportional to $m_{3/2}$ without a corresponding goldstino coupling.

In addition to the tree-level terms in \Eq{eq:SUGRAlagrange}, there is a contribution to the Lagrangian coming from anomaly matching.  Before introducing (and gauge fixing) $\bPhi$, conformal SUGRA contained a non-anomalous $U(1)_R$ gauge symmetry with gauge field $b_\mu$, so the corresponding global $U(1)_R$ must also be non-anomalous.  Under this $U(1)_R$, $\bPhi$ (which we have yet to gauge fix) has $R$-charge $2/3$ and matter fields have $R$-charge 0.  Since chiral fermions have $R$-charge $-1$ and gauginos have $R$-charge $+1$, the gauge kinetic function for each gauge field must contain
\be
\label{eq:startingAnomaly}
 \boldf_{ab} \supset  \delta_{ab} \, \frac{g_a^2}{4 \pi^2} \left(\frac{3T_R - 3 T_G}{2} \right) \log \bPhi,
\ee
such that these anomalies can be cancelled by a  $U(1)_R$ shift of $\log \bPhi$ \cite{Kaplunovsky:1994fg} (see also \Ref{Bagger:2000dh}).  Note that this is \emph{not} the familiar expression for $\bPhi$ coupling involving the beta function (see e.g.~\Ref{Randall:1998uk}).  This will arise after appropriate field redefinitions of the matter fields.

\subsection{Field Redefinitions in SUGRA}

The Lagrangian shift in \Eq{eq:SUSYshift} appears for any field rescaling of chiral multiplets, including rescalings involving the conformal compensator.  With the improved gauge fixing, there is no mixing between matter multiplets and the gravity multiplet, and this lack of mixing persists (at least at one loop) after field rescalings.\footnote{This rescaling does induce a gravitational anomaly term, but this is irrelevant for our present purposes.}  In addition to the appearance of \Eq{eq:startingAnomaly}, the main difference between K\"ahler mediation in global SUSY and full anomaly mediation in SUGRA is that $\bK$ in \Eq{eq:globalKahler} is replaced by $\bPhi^\dagger \bPhi \, \bOmega$, with
\be
\bOmega \equiv -3 e^{-\bK/3}.
\ee

We can now use the same fields manipulation as in \Sec{sec:anomalyInGlobal}, treating $\bPhi$ as one of the SUSY-breaking fields.  First, to remove linear couplings to the conformal compensator, we can perform the field redefinition
\be
\label{eq:Q2QoverPhi}
\bQ^i \rightarrow \frac{\bQ^i}{\bPhi}.
\ee
Combined with \Eq{eq:startingAnomaly}, this leads to the familiar anomaly-mediated term 
\be
\label{eq:familiarAnomaly}
\delta \mathcal{L} = - \frac{g^2}{16\pi^2} \int d^2 \theta \left(\frac{3T_G - T_R}{2} \right) \log \bPhi  {\bW}^{a \alpha} {\bW}^a_\alpha,
\ee
which is proportional to the beta function $b_0 \equiv 3T_G - T_R$ as expected.  To remove linear couplings to SUSY-breaking fields in $\bOmega$, we use \Eq{eq:generalRedef}, replacing $\bK$ with $\bOmega$  
\be
\frac{1}{d_R} (\log \det \bOmega|_R'' ) \Rightarrow - \frac{1}{3} \bK + \frac{1}{d_R} (\log \det \bK|_R'').
\ee
Here, we have used the fact that for unbroken gauge symmetries, the vev of $K_i$ (and of any derivatives of $K_i$ with respect to the SUSY-breaking fields) is zero for charged fields $\bQ^i$.  Combined with \Eq{eq:familiarAnomaly}, we arrive at the final anomaly-mediated expression\footnote{As discussed in \App{app:fullmass}, there is an additional anomaly-mediated contribution arising from rescaling gauge multiplets from a holomorphic basis to a canonical basis.  This effect is not captured by \Ref{Bagger:1999rd} since it requires direct couplings between SUSY breaking and gauginos, but it does appear in \Ref{Conlon:2010qy}.}
\be
\label{eq:finalAnomaly}
\delta \mathcal{L} = - \int d^2 \theta \, \frac{g^2}{16\pi^2}   \left(\left(\frac{3T_G - T_R}{2} \right) \log \bPhi +  \frac{\overline{D}^2 D^2}{16 \, \square}\left[- \frac{T_R}{3} \bK + \frac{T_R}{d_R} (\log \det \bK|_R'')  \right]   \right)  {\bW}^{a \alpha} {\bW}^a_\alpha.
\ee
Using the improved gauge fixing, anomaly mediation in SUGRA has essentially the same origin as K\"ahler mediation in global SUSY, arising from performing anomalous chiral rescalings to remove linear couplings to SUSY breaking in the K\"ahler potential.

As emphasized in \Ref{Bagger:1999rd}, anomaly mediation is associated with three different anomalies---a super-Weyl anomaly, a K\"ahler anomaly, and a sigma-model anomaly---corresponding to the three terms in \Eq{eq:finalAnomaly}.  In our rescaling procedure, the K\"ahler and sigma-model anomalies in SUGRA have a common origin, and arise from taking the global sigma-model anomaly involving the K\"ahler potential $\bK$ and replacing it with an ``effective'' K\"ahler potential $\bOmega$.  In this way, the  K\"ahler anomaly should be regarded as a $1/\MPl$ correction to the sigma-model anomaly.  The super-Weyl anomaly is truly a SUGRA effect, and depends crucially on the fact that prior to gauge fixing, there was an anomaly-free global $U(1)_R$ symmetry.\footnote{As a side note, the derivation of anomaly mediation in \Ref{Dine:2007me} focused only on an Abelian gauge theory, so it does not capture the $T_G$ dependence in non-Abelian theories which arises from \Eq{eq:startingAnomaly}.}

\subsection{Soft Masses and Gaugino Couplings}

Before expanding \Eq{eq:finalAnomaly} in components, there is no apparent difference between gravitino mediation and K\"ahler mediation.  This difference only becomes visible after identifying the gaugino soft masses and corresponding gaugino couplings in \Eq{eq:genericGoldstinoCouplings}, repeated for convenience:
\be
\tag{\ref{eq:genericGoldstinoCouplings}}
\mathcal{L} \supset - \frac{1}{2}m_{\lambda} \lambda_a \lambda^a + \frac{c_\lambda}{\sqrt{2} F_{\rm eff}} \lambda_a \sigma^{\mu \nu} \gold F_{\mu \nu}^a.
\ee

The gaugino mass from expanding \Eq{eq:finalAnomaly} is
\be
m_\lambda = -\frac{g^2}{16\pi^2} \left( (3T_G - T_R) \left(m_{3/2} +\frac{K_i F^i}{3} \right) - 2 \frac{ T_R}{3} K_i F^i + 2 \frac{ T_R}{d_R} (\log \det K|_R'')_i F^i  \right).
\ee
Note that both the super-Weyl and K\"ahler anomaly pieces have contributions proportional to $K_i F^i$, and we have used the fact that $\vev{F_\phi} = m_{3/2}$ in the improved gauge fixing from \Eq{eq:improvedGauge}.     We can extract the goldstino coupling $c_\lambda$ from \Eq{eq:finalAnomaly}, using \Eq{eq:goldstinoMode} to identify the goldstino direction:
\be
c_\lambda = -\frac{g^2}{16\pi^2} \left( (3T_G - T_R) \frac{K_i F^i}{3} - 2 \frac{T_R}{3} K_i F^i + 2 \frac{T_R}{d_R} (\log \det K|_R'')_i F^i  \right).
\ee
Crucially, $c_\lambda$ differs from $m_\lambda$ by terms proportional to $m_{3/2}$, owing to the fact that the $(1 + \theta^2 m_{3/2})$ piece of $\bPhi$ has a SUSY-breaking $F$-component without a corresponding goldstino components.  These terms are summarized in \Tab{tab:summary}.

We can rewrite the gaugino mass and goldstino coupling in the following suggestive way:
\be
\label{eq:finalAnswer}
m_\lambda = m_{\rm AdS} + c_\lambda,
\ee
where 
\begin{align}
m_{\rm AdS} &= -\frac{g^2}{16 \pi^2} \left(m_{3/2} (3 T_G - T_R) \right), \label{eq:mAdS}\\
c_\lambda &= - \frac{g^2}{16 \pi^2} \left(\frac{K_i F^i}{3} (3T_G - 3T_R) + 2 \frac{T_R}{d_R}(\log \det K|_R'')_{,i}F^i  \right).
\end{align}
This is the primary result of this paper.  Here, $m_{\rm AdS}$ is the gaugino mass splitting from the bulk of SUSY AdS space (derived in \Ref{Gripaios:2008rg} and discussed further in \Sec{sec:goldstinoequivalence}), and gives rise to a gravitino-mediated soft mass with no associated goldstino coupling.  The remaining part of anomaly mediation $c_\lambda$ is K\"ahler mediation, which generalizes the global SUSY results from \Sec{sec:anomalyInGlobal}.  As advertised, $c_\lambda$ is an effective measure of sequestering---in particular, sequestering of visible sector gauginos from the goldstino---and the limit $c_\lambda = 0$ corresponds to pure gravitino mediation.

\section{Alternative Descriptions}
\label{sec:SUGRAmultiplet}

Having seen the two faces of anomaly mediated in the conformal compensator formalism, it is worth repeating the calculation in the (perhaps) more familiar language of \Ref{Bagger:1999rd}.  We  first rederive \Eq{eq:finalAnswer} in components, and then explain the connection to the AdS supercurrent and the goldstino equivalence theorem. 

\subsection{Anomaly Mediation in Components}

As shown in \Refs{Buchbinder:1988yu,LopesCardoso:1993sq,Bagger:1999rd}, after lifting the super-Weyl, K\"ahler, and sigma-model anomalies to superspace, the 1PI effective action contains
\be
\label{eq:supergravityFrameAnomaly}
\mathcal{L}_{\rm SF} \supset -\frac{g^2}{256 \pi^2} \int d^2 \Theta \, 2 \mathcal{E} \, \bW^\alpha \bW_\alpha \bC,
\ee
where for convenience, we have defined a chiral superfield $\bC$ as
\be
\label{eq:superfieldC}
\bC = \frac{1}{\square} \left(\overline{\mathcal{D}}^2 - 8 \bR  \right) \left[ 4 (T_R - 3 T_G) \bR^\dagger - \frac{1}{3} T_R \mathcal{D}^2 \bK + \frac{T_R}{d_R} \mathcal{D}^2 \log \det \bK|_R''   \right],
\ee
where $\bR$ is the curvature superfield.  This expression is valid in ``supergravity frame'' where the Einstein-Hilbert term has the non-canonical normalization $e^{-K/3} R_{\rm EH}$.\footnote{It may be confusing that \Eq{eq:supergravityFrameAnomaly} is only a function of the K\"ahler potential $K$ and not the K\"ahler invariant $G \equiv K + \log W + \log W^*$.  Because of the K\"ahler anomaly, there is a physical distinction between the superpotential $W$ and the holomorphic terms in $K$.  See \Refs{Bagger:1999rd,Bagger:2000dh}.}  By taking the lowest component of $\bC$ in \App{app:lowestC}, we recover (non-local) terms in the Lagrangian that express the three anomalies.

In order to derive physical couplings and masses from the other components of $\bC$, we need to transform to ``Einstein frame'' where the graviton (and gravitino) have canonical kinetic terms.  This can be accomplished by performing the field redefinitions \cite{Bagger:1999rd, wess1992supersymmetry, LeDu:1998ne}
\begin{align}
e_c^\mu &\rightarrow e^{-K/6} e_c^\mu, \\
\psi_\mu &\rightarrow e^{+K/12} (\psi_\mu + i \sqrt{2} \sigma_\mu \overline{\chi}^\i K_\i / 6),  \label{eq:shiftgravitino}  \\
M^* &\rightarrow e^{-K/6}(M^* - F^i K_i), \label{eq:shiftM} \\
\lambda^a &\rightarrow e^{-K/4} \lambda^a, \\
\chi^i &\rightarrow e^{-K/12} \chi^i,\\
F^i & \rightarrow e^{-K/6} F^i, \\
b_\mu & \rightarrow b_\mu + \frac{i}{2} (K_i \partial_\mu A^i - K_\i \partial_\mu A^{* \i}) .
\end{align}
Note that the gravitino $\psi_\mu$, scalar auxiliary field $M^*$, and vector auxiliary field $b_\mu$ transform inhomogeneously under this redefinition.\footnote{Indeed, the improved gauge fixing of \Ref{Cheung:2011jp} was designed to avoid having to perform such transformations.}  With this field redefinition and adjusting the cosmological constant to zero, the scalar auxiliary vev is
\be
\vev{M^*} = -3 m_{3/2},
\ee
analogous to \Eq{eq:Fphivev}.

After performing the field redefinitions, the pertinent components of $\bC$ are 
\begin{align}
e^{K/12} \, \D_\alpha \bC | & =   \frac{16}{3 \sqrt{2}} (3 T_G - T_R) K_i \chi^i_\alpha - \frac{32}{3 \sqrt{2}} T_R \left<K_i\right> \chi^i_\alpha + \frac{32}{\sqrt{2}} \frac{T_R}{d_R} \left<(\log \det K|^{''}_{R})_i\right> \chi^i_\alpha\nonumber \\
& \quad \; - \frac{32}{3 \Box} (3 T_G - T_R) (i \sigma^\mu \sigmabar^{\nu \rho} \D_\mu \D_\nu \psi^\dagger_\rho)_\alpha+ \cdots, \label{eq:Ccoupling} \\
e^{K/6} \, \D^2 \bC | & = \frac{32}{3} (3 T_G - T_R) (- 3 m_{3/2} - F^i K_i) + \frac{64}{3} T_R K_i F^i \nonumber \\
& \quad  \; - 64 \frac{T_R}{d_R} (\log \det K|^{''}_R)_i F^i+ \cdots. \label{eq:Cmass} 
\end{align}
Here, it is understood that we have shifted all fields such that their vevs are zero and any expressions contained in angle brackets above are purely c-numbers.  The ellipses represent omitted terms that do not correspond to any local terms in the resultant Lagrangian, but are necessary to maintain SUSY in the 1PI action.  The gravitino coupling in the last term of \Eq{eq:Ccoupling} will be important in \Sec{sec:goldstinoequivalence} below.

The $\Theta^2$ component of $\bC$ yields the gaugino soft mass, and the $\Theta$ component of $\bC$ yields the gauge boson-gaugino-goldstino coupling.  We can now derive \Eq{eq:genericGoldstinoCouplings}, after identifying the goldstino mode through \Eq{eq:goldstinoMode}, and we recover the same answer as \Eq{eq:finalAnswer}:
\be
\mathcal{L} \supset - \frac{1}{2}\left(m_{\rm AdS} + c_\lambda \right) \lambda_a \lambda^a + \frac{c_\lambda}{\sqrt{2} F_{\rm eff}} \lambda_a \sigma^{\mu \nu} \gold F_{\mu \nu}^a,
\ee
with
\begin{align}
m_{\rm AdS} &=  -\frac{g^2}{16 \pi^2} \left(m_{3/2} (3 T_G - T_R) \right),\\
c_\lambda &=  -\frac{g^2}{16 \pi^2} \left(\frac{K_i F^i}{3}(3 T_G - 3T_R) + 2 \frac{T_R}{d_R}(\log \det K|_R'')_{,i}F^i  \right).
\end{align}
Because of the gravitino shift in \Eq{eq:shiftgravitino} and the auxiliary field shift in \Eq{eq:shiftM}, the super-Weyl anomaly contributes to both gravitino mediation and K\"ahler mediation.  Again, we see that gravitino mediation is physically distinct from K\"ahler mediation by the absence of goldstino couplings.

\subsection{Supercurrent Conservation and Goldstino Equivalence}
\label{sec:goldstinoequivalence}

The fact that gravitino mediation gives rise to gaugino soft masses without corresponding goldstino couplings is perhaps confusing from the point of view of the goldstino equivalence theorem \cite{Casalbuoni:1988qd, Casalbuoni:1988kv}.  However, we will see that this is necessitated by conservation of the AdS supercurrent.

The goldstino equivalence theorem states that at energies well above the gravitino mass $m_{3/2}$, the couplings of longitudinal gravitinos can be described by the (eaten) goldstino mode.  In global SUSY, linear couplings of the goldstino are fixed by conservation of the (flat space) supercurrent 
\be
\label{eq:supercurrent}
\mathcal{L} = \frac{1}{\sqrt{2} F_{\rm eff}} \partial_\mu \gold j_{\rm flat}^\mu.
\ee
The part of the supercurrent that depends on the gauge boson and gaugino is
\be
j_{\rm flat}^\mu \supset  - \frac{i}{2} \sigma^\nu \overline{\sigma}^\rho \sigma^\mu \overline{\lambda}_a F^a_{\nu \rho}.
\ee
Using the gaugino equation of motion (assuming a massless gauge boson for simplicity), this gives rise to the interaction 
\be
\label{eq:flatspaceGaugeGauginoGoldstino}
\text{Flat space:} \quad \frac{m_\lambda}{\sqrt{2} F_{\rm eff}} \lambda_a \sigma^{\mu \nu} \gold F^a_{\mu \nu},
\ee
where $m_\lambda$ is the physical gaugino mass.  In flat space, therefore, $c_\lambda$ must equal $m_\lambda$, and there can be no contribution from gravitino mediation.

The resolution to this apparent paradox is that the gravitino mediation arises from uplifting an AdS SUSY vacuum to SUSY-breaking Minkowski space, so we should really be testing the goldstino equivalence theorem for (rigid) AdS space \cite{Keck:1974se,Zumino:1977av,Ivanov:1979ft,Ivanov:1980vb}.\footnote{Rigid AdS corresponds to the limit $\MPl \rightarrow \infty$ leaving the AdS curvature fixed.  This limit maintains couplings proportional to $m_{3/2}$ despite the fact that the gravitino itself is decoupled.}  Indeed, the last term in \Eq{eq:Ccoupling} contains an additional coupling to the gravitino, which contributes to the (AdS) supercurrent.\footnote{Strictly speaking, this term contributes only to the bulk AdS$_4$ supercurrent, as there is an additional boundary term that compensates to allow massless gauginos in SUSY AdS$_4$ \cite{Gripaios:2008rg}.  After lifting AdS space to flat space, this boundary term becomes irrelevant.}
 In principle, it should be possible to derive the one-loop AdS supercurrent directly from the SUSY algebra in AdS space, but we know of no such derivation in the literature.  Instead, we can simply extract the one-loop contribution to the supercurrent by recalling that the gravitino couples linearly to the (AdS) supercurrent as
\be
\label{eq:supercurrentSUGRA}
\mathcal{L} = - \frac{1}{2 M_{\rm Pl}} \psi_\mu j_{\rm AdS}^\mu \rightarrow \frac{1}{\sqrt{2} F_{\rm eff}} \partial_\mu \gold j_{\rm AdS}^\mu.
\ee
In this last step, we have identified the goldstino direction via \cite{wess1992supersymmetry} 
\be
\label{eq:goldstinodirection}
\psi_\mu \rightarrow  - \sqrt{\frac{2}{3}} m_{3/2}^{-1} \partial_\mu \gold - \frac{i}{\sqrt{6}} \sigma_\mu \goldbar, \qquad m_{3/2} = \frac{F_{\rm eff}}{\sqrt{3} M_{\rm Pl}},
\ee
and dropped the term proportional to $\goldbar \, \sigmabar_\mu j^\mu_{\rm AdS}/\MPl$ since it does not contain a gauge boson-gaugino-goldstino coupling.  

We see that \Eq{eq:Ccoupling} contains a linear (non-local) coupling to the gravitino, and thus an additional (local) coupling to the goldstino
\be
\left( \frac{g^2}{16 \pi^2} m_{3/2} (3 T_G - T_R)\right) \frac{1}{\sqrt{2} F_{\rm eff}} \lambda_a \sigma^{\mu \nu} \gold F^a_{\mu \nu}.
\ee
We recognize the term in parentheses as $-m_{\rm AdS}$ from \Eq{eq:mAdS}.  Combining with \Eq{eq:flatspaceGaugeGauginoGoldstino}, the full goldstino coupling in SUGRA is
\be
\text{AdS space:} \quad  \frac{m_\lambda - m_{\rm AdS}}{\sqrt{2} F_{\rm eff}}  \lambda \sigma^{\mu \nu} \gold F_{\mu \nu},
\ee
in perfect agreement with \Eq{eq:finalAnswer}.  Thus, the goldstino equivalence theorem holds even in the presence of gravitino mediation, albeit with the AdS supercurrent.  This is as we anticipated, since particles and sparticles have SUSY mass splittings in the bulk of AdS space, so ``soft masses'' arising from $m_{\rm AdS}$ should not have an associated goldstino coupling. We could alternatively derive the same effect in unitary gauge for the gravitino by realizing that the last term in \Eq{eq:Ccoupling} modifies longitudinal gravitino interactions by an amount proportional to $m_{\rm AdS}/m_{3/2}$.

\section{Discussion}
\label{sec:conclusions}

In this paper, we have shown that anomaly mediation consists of two physically distinct phenomena, which can be distinguished by their associated goldstino couplings.  Gravitino mediation (i.e.\ traditional anomaly mediation) is familiar from the phenomenology literature, but it has the counter-intuitive feature that it has no associated goldstino coupling.  Indeed, the difference $m_\lambda - c_\lambda = m_{\rm AdS}$ is a physical way to measure gravitino mediation, and $c_\lambda$ characterizes the degree of sequestering between the visible sector and the goldstino.  K\"ahler mediation simply arises from linear couplings of SUSY-breaking fields in the K\"ahler potential, and appears in both global and local SUSY.  The soft masses and goldstino couplings from K\"ahler mediation satisfy the (flat space) goldstino equivalence theorem.

While these two faces of anomaly mediation can be understood directly in SUGRA component fields as in \Sec{sec:SUGRAmultiplet}, the physics is more transparent using the improved gauge fixing of \Ref{Cheung:2011jp}.  In this gauge, it is obvious why soft masses proportional to $m_{3/2}$ do not have any associated goldstino couplings, since the conformal compensator $\bPhi$ contains a piece $(1 + \theta^2 m_{3/2})$ with no fermionic component.  It is also obvious that the super-Weyl anomaly contributes both to gravitino mediation and to K\"ahler mediation.  For deriving K\"ahler mediation in SUGRA, it is convenient that the K\"ahler and sigma-model anomalies are tied together into a single $\bOmega$ function.

As previewed in the introduction, the case of scalar soft masses is more subtle, and we leave a detailed study to future work \cite{scalars:2012}.   For gravitino mediation, conservation of the AdS$_4$ supercurrent must hold to all loop orders, such that any soft mass proportional to the AdS curvature will have no associated goldstino coupling.  However, tree-level tachyonic scalars masses given by $-2m_{3/2}^2$ must be compensated by SUSY breaking to have a stable theory in flat space.  This tachyonic piece is in addition to the well-known two-loop anomaly-mediated soft masses, so even in sequestered theories, there will be irreducible (but unambiguous) couplings between matter multiplets and the goldstino.  Since anomaly mediation can be alternatively derived using Pauli-Villars regulating fields \cite{Giudice:1998xp,Gaillard:1999yb,Gaillard:2000fk}, we should find that the soft masses and goldstino couplings of the regulators are precisely those necessary to maintain the gravitino/K\"ahler mediation distinction in the regulated theory.

We have emphasized the fact that a gaugino soft mass can appear with no associated goldstino couplings in the case of strict sequestering, which yields pure gravitino mediation.  Interestingly, there are also reversed cases where a goldstino coupling is present with no associated gaugino mass.  Famously, anomaly mediation is absent in no-scale SUSY breaking (and suppressed in almost-no-scale models) \cite{Luty:2002hj}.  Also, theories with extra-dimensional warping can have suppressed anomaly mediation \cite{Luty:2002ff}.  However, these arise from a cancellation between gravitino mediation and K\"ahler mediation (through moduli $F$-components), and thus goldstino couplings are still present even when there are no anomaly-mediated soft masses.  This bizarre result is nevertheless required by conservation of the AdS supercurrent, and emphasizes the fact that the underlying symmetry structure of our universe is not just SUSY, but SUSY in AdS space.

\begin{acknowledgments}

We benefitted from conversations with Allan Adams, Jonathan Bagger, Clifford Cheung, Joseph Conlon, Daniel Freedman, David E. Kaplan, Markus Luty, Aaron Pierce, Erich Poppitz, and Raman Sundrum.  This work was supported by the U.S. Department of Energy (DOE) under cooperative research agreement DE-FG02-05ER-41360.  J.T. is also supported by the DOE under the Early Career research program DE-FG02-11ER-41741.

\end{acknowledgments}

\appendix

\section{The Fourth Anomaly in Anomaly Mediation}
\label{app:fullmass}

As mentioned in \Tab{tab:summary} and footnote \ref{foot:conlon}, there is a fourth anomaly which can contribute to the gaugino mass, though it is not so important for phenomenology since it requires direct couplings of SUSY breaking to the gauginos at tree-level.  It was first pointed out in \Ref{Conlon:2010qy} in a string theory context.  For completeness, we derive in this appendix the extra contribution within our framework, and we show that the associated goldstino coupling respects (flat space) supercurrent conservation. 

Following the notation in \Ref{ArkaniHamed:1998kj}, the Yang-Mills term in a SUSY gauge theory is
\be
\mathcal{L} \supset  \frac{1}{2} \int d^2 \theta  \, {\bs S} \,   {\bW}^{a \alpha} {\bW}^a_\alpha ,
\label{eq:Shol}
\ee
where ${\bs S}$ is the holomorphic gauge coupling. The superfield ${\bs S}$ is chiral and does not run beyond one-loop in perturbation theory.  However, the component fields of the gauge multiplet appearing in \Eq{eq:Shol} are not canonically normalized.  In order to go to a canonically-normalized basis, we need to perform an anomalous rescaling of the gauge multiplet.  This will induce an additional anomaly-mediated contribution to the gaugino mass.  

As shown in \Ref{ArkaniHamed:1998kj}, the effects of this rescaling are encoded in the real vector superfield ${\bs R}$ (not to be confused with the curvature superfield), given by\footnote{The elided terms include the sigma-model anomaly term already contained in \Eq{eq:generalRedef}.}
\be
{\bs R} \equiv \left({\bs S} + {\bs S}^\dag \right) + \frac{T_G}{8 \pi^2} \log\left[ {\bs S} + {\bs S}^\dag \right] + \ldots .
\label{eq:RvsS}
\ee
The physical meaning of the components of ${\bs R}$ can be identified from the 1PI effective action
\be
\mathcal{L}_{\rm 1PI} = \int d^4 \theta \, \bR \,  {\bW}^{a \alpha} \frac{D^2}{-8 \square} {\bW}^a_\alpha + \textrm{h.c.}
\ee
The lowest component of ${\bs R}$ defines the canonical gauge coupling, and the $\theta^2$ component is related to the physical gaugino mass, via
\be
\frac{1}{g^2} = \bR\bigr|_{\theta^0}, \qquad m_\lambda = -  \log {\bs R} \bigr|_{\theta^2}.
\ee
The physical gaugino-gauge boson vertex is determined by 
\be
\label{eq:Rvertex}
\mathcal{L} \supset - \frac{1}{2} \lambda_a \sigma^{\mu \nu} F_{\mu \nu}^a \log {\bs R}\bigr|_{\theta} .
\ee

If ${\bs S}$ has a $\theta^2$ component at tree-level, then there is an extra contribution to the gaugino mass and goldstino coupling from the  second term in \Eq{eq:RvsS}, in addition to the expected tree-level gaugino mass and goldstino coupling from the first term.  This additional piece due to the anomalous rescaling of the gauge multiplet is
\be
\Delta m_\lambda = - \frac{g^2  \, T_G}{8 \pi^2} \, F^i \, \partial_i \log S .
\ee
We can also read off the associated goldstino coupling from \Eqs{eq:RvsS}{eq:Rvertex}, after identifying the goldstino direction through \Eq{eq:goldstinoMode}. This gives an additional goldstino coupling
\be
\Delta c_\lambda = \Delta m_\lambda
\ee
in the notation of \Eq{eq:genericGoldstinoCouplings}, consistent with (flat space) supercurrent conservation.

\section{General Chiral Field Redefinitions}
\label{app:generalKahler}

In order to derive \Eq{eq:generalRedef}, we want to find a field redefinition on our (charged) matter superfields of the form 
\be
\label{eq:chiralredef}
\bQ^i \rightarrow e^{\bs \alpha^i} \bQ^i 
\ee
that removes all chiral couplings of the $\bQ^i$ to the SUSY-breaking fields $\bX^i$, while preserving the canonical normalization of all kinetic terms.  Explicitly, we want that after this field redefinition,
\be
\label{eq:fieldredefgoal}
\left<\bK_{i \j} \right> = \delta_{i \j}, \quad \vev{\bK_{i \j \ell}} = 0,
\ee
where exactly one of the indices on the latter corresponds to a SUSY-breaking field.

Assuming we have shifted away all vevs of our scalar fields, the most general K\"ahler potential for charged matter can be written as
\be
\label{eq:generalKahler}
\bK = \bQ^i \bQ^{\dagger \j} \delta_{i \j} + A_{i \j \ell} \bQ^i \bQ^{\dagger \j} \bX^\ell + \textrm{h.c.} + \cdots,
\ee
where we have omitted any terms that have no impact on \Eq{eq:fieldredefgoal} and rotated and rescaled the matter fields to have canonical kinetic terms.   The linear couplings to $\bX^\ell$ can be removed by the field redefinition 
\be
\bQ^i \rightarrow e^{- \frac{\overline{\D}^2 \D^2}{16 \Box} (\log \bK'')_{k \i} } \bQ^k = (\delta_{k \i} - A_{k \i \ell} \bX^\ell + \cdots ) \bQ^k
\ee
with $\bK''$ being the K\"ahler metric.  This redefinition induces the anomaly term
\begin{eqnarray*}
\delta \mathcal{L} & = & \sum_i \int d^2 \theta \, \frac{g^2}{16 \pi^2} T_{R_i} \left(-  \frac{\overline{D}^2 D^2}{16 \Box} (\log \bK'')_{i \i} \right) \bW^{a \alpha} \bW^a_\alpha \\
& = & \sum_R - \frac{g^2}{16 \pi^2} \int d^2 \theta \, \frac{T_R}{d_R} \frac{\overline{D}^2 D^2}{16 \Box} \log \det \bK|_R'' \, \bW^{a \alpha} \bW^a_\alpha.
\end{eqnarray*}
The sum in the last line is now over the matter representations $R$.

\section{Non-Local Anomaly Terms}
\label{app:lowestC}

The lowest component of the superfield $\bC$ in \Eq{eq:superfieldC} yields (non-local) terms in the Lagrangian that express the three anomalies of the theory.  In supergravity frame, we have explicitly
\begin{align}
\bC | & = \frac{1}{\Box} \left[\frac{8}{3} (T_R - 3 T_G ) \left(- \frac{1}{2} \mathcal{R}  +i \partial_\mu b^\mu    \right)    - \frac{16}{3} T_R (K_i \Box A^i + K_{i j} \D_\mu A^i \D^\mu A^j ) \right. \nonumber \\
& \quad \quad \; \left.  + 16 \frac{T_R}{d_R}  \left( (\log \det K|^{''}_R )_i \Box A^i + (\log \det K|^{''}_R )_{i j} \D_\mu A^i \D^\mu A^j  \right) + \cdots \right] \label{eq:lowestC}.
\end{align}
For example, the super-Weyl anomaly (or more accurately, the $U(1)_R$ anomaly \cite{Bagger:1999rd}) is expressed via
\be
\mathcal{L} \supset \frac{g^2}{96 \pi^2} (3 T_G - T_R) \frac{\partial_\rho b^\rho}{\square}F_{\mu\nu}\widetilde{F}^{\mu\nu},
\ee
where $b_\rho$ is the vector auxiliary field which shifts as $b_\rho \rightarrow b_\rho + \partial_\rho \alpha$ under a $U(1)_R$ transformation.  Rearranging \Eq{eq:lowestC}, the K\"ahler anomaly and sigma-model anomaly are similarly expressed via the K\"ahler connection and sigma-model connection \cite{Bagger:1999rd}:  
\begin{align}
\mathcal{L} & \supset  - \frac{g^2}{96 \pi^2} T_R \frac{\partial_\rho (i K_i \partial^\rho A^i - i K_\i \partial^\rho A^{* \i} ) }{\Box} F_{\mu \nu} \widetilde{F}^{\mu \nu}, \\
\mathcal{L}  & \supset  \frac{g^2}{32 \pi^2} \frac{T_R}{d_R} \frac{\partial_\rho (i (\log \det K|^{''}_{R})_i \partial^\rho A^i - i (\log \det K|^{''}_{R})_\i \partial^\rho A^{* \i})}{\Box} F_{\mu \nu} \widetilde{F}^{\mu \nu}.
\end{align}

\bibliography{AnomalyMed}
\bibliographystyle{JHEP}

\end{document}